\documentclass{article}

\usepackage{arxiv}

\usepackage{graphicx} 
\usepackage[T1]{fontenc}
\usepackage{microtype}
\usepackage{graphicx}
\usepackage{booktabs}
\usepackage{caption}
\usepackage{subcaption}
\usepackage{hyperref}
\usepackage{amssymb}
\usepackage{mathtools}
\usepackage{amsthm}
\usepackage{xcolor}
\usepackage{multicol}
\usepackage[natbibapa]{apacite}
\newtheorem{assumption}{Assumption}
\newtheorem{presumption}{Presumption}

\renewcommand{\shorttitle}{The History and Risks of RL and HF}

\title{The History and Risks of \\Reinforcement Learning and Human Feedback}

\begin{document}
\author{
  Nathan Lambert\\
  Allen Institute for AI \\
  Berkeley, CA, USA\\
  \texttt{nathanl@allenai.org} \\
   \And
  Thomas Krendl Gilbert \\
  The New York Academy of Sciences \\
  New York, NY, USA \\
  \texttt{tgilbert@nyas.org} \\
     \And
  Tom Zick \\
  Berkman Klein Center, Harvard Law \\
  Cambridge, MA, USA \\
  \texttt{tzick@jd24.law.harvard.edu} \\
}

\date{\today}

\maketitle

\begin{abstract}
Reinforcement learning from human feedback (RLHF) has emerged as a powerful technique to make large language models (LLMs) easier to use and more effective.
A core piece of the RLHF process is the training and utilization of a model of human preferences that acts as a reward function for optimization.
This approach, which operates at the intersection of many stakeholders and academic disciplines, remains poorly understood.
RLHF reward models are often cited as being central to achieving performance, yet very few descriptors of capabilities, evaluations, training methods, or open-source models exist.
Given this lack of information, further study and transparency is needed for learned RLHF reward models.
In this paper, we illustrate the complex history of optimizing preferences, and articulate lines of inquiry to understand the sociotechnical context of reward models. 
In particular, we highlight the ontological differences between \textit{costs}, \textit{rewards}, and \textit{preferences} at stake in RLHF's foundations, related methodological tensions, and possible research directions to improve general understanding of how reward models function.
\end{abstract}

\section{Introduction}
Learning from human feedback has become incredibly popular due to the success of large language models (LLMs) such as OpenAI's ChatGPT~\citep{ChatGPT} and Anthropic's Claude~\citep{bai2022training}, which are heavily dependent on human labeled data.
These models make use of reinforcement learning from human feedback (RLHF), a technique designed to integrate human preferences where writing an explicit reward function is otherwise challenging~\citep{christiano2017deep}.
In the context of language models, RLHF proceeds as follows: first, a reward model is independently trained on aggregate pairwise preferences from many crowdworkers to rate any piece of text; second, the language model is optimized with an RL optimizer~\citep{ouyang2022training, bai2022training, touvron2023llama}.
The final language model is often subject to heavy scrutiny both internally and, more recently, externally through events like DEFCON Red Teaming Village~\citep{defcon} or coordinated adversarial attacks~\cite{zou2023universal}. 
The same cannot be said for the intermediate reward model. 
Historically, reward models have not been released as open-source or evaluated rigorously, obscuring from scrutiny the process through which values are actively encoded into the system. 
This paper illustrates why reward models are central to understanding the long-term impacts of RLHF, drawing from the rich history, discourse, and tension around how to best quantify human values.

RLHF is the intellectual culmination of several distinct domains. 
The optimization stack of RLHF is borrowed from control theory, a domain in which there are ground truths and reward functions can have an clear notion of success. 
The primary risk of learning human preferences for LLMs comes through the domain shift from control to language.
In language, notions of values are more computationally complex or fundamentally vague~\cite{dobbe2021hard} in relation to their control counterparts, but the optimization stack nevertheless remains similar. 
Reinforcement learning is broadly the field of study of sequential decision making, which is built on a substantial literature comprising cognitive biology, optimal control, behavioral economics, and other fields~\citep{sutton2018reinforcement}.
RLHF combines the social challenges of human data with the techniques of RL --- a field with a long history of numerical complexity. 

Despite the maturity of the domains it draws on, grounding and investigating risks of RLHF requires the development of new tools and research methods. 
In particular, vestigial assumptions inherited from earlier technologies can re-surface as blind spots in the modern RLHF paradigm.
Tracing the history of RL and RLHF as technologies allows us to identify these assumptions and where they matter in particular systems.  
This paper attempts to provide an exposition of some of this historical context, and to highlight specific sociotechnical areas of opportunity within the reward model specification, beyond the challenges proposed in recent literature~\citep{casper2023open}.
We study the histories of quantification of human preferences and reinforcement learning algorithms, from \textit{Port-Royal Logic} and Bentham to Markov Decision Processes and Bellman, to highlight potential shortcomings of learning models of human preferences.
An initial concern that has been raised with the current deployments of LLMs is the limitations of working with aggregate human data, raising questions as to whose values the model is encoding and prioritizing.
Moving beyond this, we study how structural optimization and deployment decisions can impact downstream users.

Given the nuance around modeling human preferences, we refer to these artifacts as reward models of preference, or \textit{reward models} for short, to better match their usage as an optimization target for reinforcement learning algorithms rather than an accurate representation of human values.
In order to broaden the scope of study around these reward models, we make the following contributions:
\begin{itemize}
    \item \textbf{Trace the complex intellectual history of RLHF} to illustrate the potential ill-posed assumptions popularized within RLHF. 
    In Sec.~\ref{sec:origins}, we explain the evolution of RL with the history of rational agents and human preferences. 
    In doing so, we distinguish sets of \textit{assumptions} (explicit premises) and \textit{presumptions} (implicit premises) made throughout the evolution of RLHF that lead to potential shortcomings of reward models.
    \item \textbf{Propose a series of questions for contemporary RLHF reward models} to increase transparency and opportunities for broader multi-stakeholder engagement in modern LLM development.
    In Sec.~\ref{sec:questions}, we break these questions down by sections of the machine learning process: data, model, and optimization, and in Sec.~\ref{sec:discussion}, we also discuss emerging issues that are not easily classified.
    \item \textbf{Discuss solutions} in Sec.~\ref{sec:sol} to measure and communicate the values and potential harms of contemporary RLHF reward models.
    We propose tools that can be used to add rigour to future empirical evaluation work.
\end{itemize}
\section{Related Works}

\subsection{Reinforcement learning from human feedback}
RLHF is a set of techniques designed to optimize machine learning models based on human feedback in order to circumvent the need to design a complex reward function. 
Early work in RLHF focused on soliciting complex behaviors from AI agents in control problems using various environments, feedback methods across trajectories or rankings, and optimizers ~\citep{christiano2017deep, wirth2017survey}.

Recently, developments in RLHF have been centered around its use with LLMs.
This branch of study originated with work exploring how technical value alignment may scale with learned reward models ~\citep{leike2018scalable}. 
Quoting \citet{leike2018scalable}:
\begin{quote}
We claim that the approach described is agnostic to the ethical paradigm, the user’s preferences, and the legal or social framework, provided we can supply enough feedback (though the preference payload might influence the amount of feedback required).
\end{quote}
The organization that builds \textit{applications} where RLHF is used bears the burden of specifying the ethics they used and answering questions about whose preferences are included and how they're weighed~\citep{prasad2018social,baum2020social}.

The development of these methods has accelerated markedly, with many variations on the methodology for integrating feedback into language models~\citep{fernandes2023bridging}.
Initial work on RLHF for LLMs utilized user choices from a batch of 4 completions~\citep{ziegler2019fine} to train a reward model across general LLM benchmarks. 
When comparing recent RLHF work to~\citet{ziegler2019fine}, group preferences were changed to pairwise preferences, and rather than general benchmarks the reward model was focused on the task of summarization~\citep{stiennon2020learning,wu2021recursively}.
Next emerged general question-answering models~\citep{ouyang2022training} and web crawling agents~\citep{nakano2021webgpt}, primarily from scaling the initial model and human datasets.
Now, RLHF is used to train general chat models across a variety of tasks~\citep{bai2022training, ChatGPT, touvron2023llama} and for specific objectives such as harm reduction~\citep{glaese2022improving} or information accuracy~\citep{menick2022teaching}, but methods for collecting the feedback data (from both humans and LLMs) are still burdened by disagreement and other technical challenges~\citep{bansal2023peering}.

\subsection{Downstream impacts of optimizing preferences}
Research venues have encouraged scientists to grapple with these questions around their work and data enrichment through humans, which are particularly relevant for techniques similar to RLHF, but there has been mixed uptake~\citep{hawkins2023ethical}.
RLHF faces many challenges with its integration of human preferences in an aggregate manner, and potential solutions involving personalization of preferences raise further questions of which values or norms are acceptable to encode in a model~\citep{kirk2023personalisation}.
Specifically, the reward models trained for RLHF are known to be over-optimized during the RL stage, where the language generations continue to shift without the reward  model indicating a higher score, without clear measurement of how downstream training signals for LLMs relate to preferences being correlated in the data~\citep{gao2022scaling}.

Training models based on human preferences also impacts how users interact with the downstream machine learning systems that refer to RLHF reward models as part of their optimization objective.
This was illustrated with the launch and widespread adoption of ChatGPT, raising questions regarding the effects of regular communication with RLHF trained LLMs, such as the downstream impact on users' moral judgements~\citep{krugel2023chatgpt} or exposure to value judgements and biases~\citep{johnson2022ghost}.
Additionally, there are open questions about the stability and robustness of RLHF-trained LLMs, with reports of RLHF models' tone shifting substantially within the course of a single conversation in potentially troubling ways~\citep{nardo2023waluigi}.

The issue of downstream model impacts is not new - for instance there is a vast prior literature on how models interface with society.
For example, user facing recommendation models have long prompted inquiry around whether agents should respond to our stated or implied preferences~\citep{milli2017should}. 
In RLHF, these concerns meet the `reward hacking' problem endemic to RL. 
Specifically, as popular models are being tuned based on user experiences, complex feedback dynamics can emerge via the combination of reward mis-specification with the power of RL optimizers~\citep{gilbert2022choices}, such as desired capabilities coming and going through repeated training.

\section{The Origins of Reward Models: Costs vs. Rewards vs. Preferences}
\label{sec:origins}

\begin{figure}[t]
    \centering
    \includegraphics[width=\linewidth]{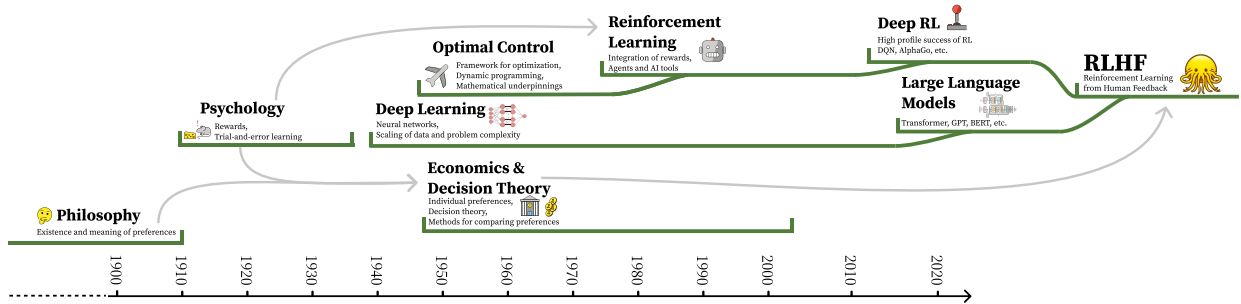}
    \caption{The timeline of the integration of various subfields into the modern version of RLHF.
    The direct links are continuous developments of specific technologies, and the arrows indicate motivations and conceptual links.}
    \label{fig:tree}
\end{figure}

\begin{figure}[t]
    \centering
    \includegraphics[width=\linewidth]{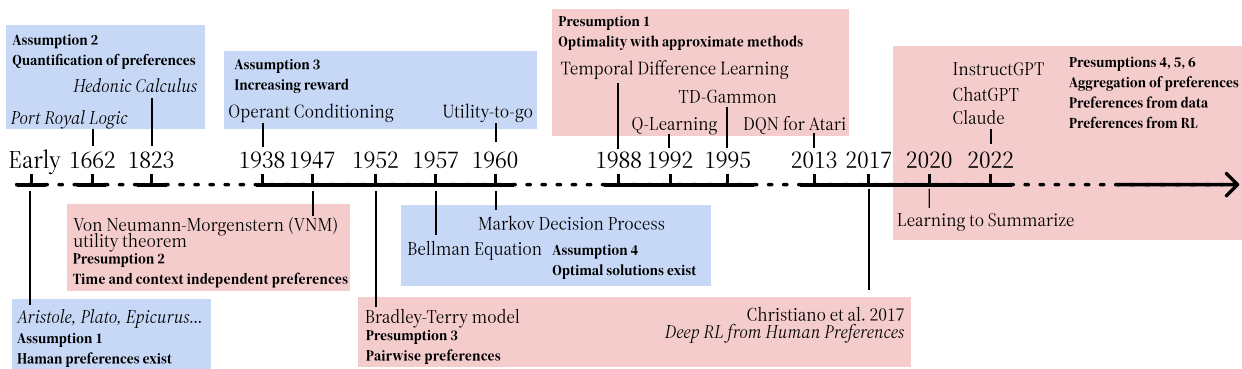}
    \caption{The history covered in Sec.~\ref{sec:origins} that creates the assumptions and presumptions central to the current deployments of RLHF.
    The assumptions indicate core theoretical foundations which RLHF builds upon, transposes, prioritizes, or defers to another development stage.
    The presumptions represent ideas and practices required to build the current renditions of the technology.}
    \label{fig:history}
\end{figure}

In this section, we break down the complex history inspiring the modern use of RLHF.
This requires investigation into the intellectual foundations of quantifying human values, reinforcement learning and optimality, as well as behavioral economics as it relates to measuring preferences.
The notion of using reinforcement learning to optimize a reward model of preferences combines the history of various once-distanced fields into an intimate optimization built on variegated assumptions about human nature.
A high level timeline illustrating the history of this foundational content is shown in Fig.~\ref{fig:tree}.
The detailed presumptions and assumptions we reference, are showcased in Fig.~\ref{fig:history}.

Our goal is to unspool the types of uncertainty that designers have grafted to system architectures at various stages of their intellectual history. 
Modern problem specifications have repeatedly stepped away from domains where optimal solutions are possible and deployed under-specified models as approximate solutions. 

Throughout, we distinguish between a series of \textit{assumptions} accepted within theoretically-grounded academic literatures, and relevant \textit{presumptions} which are common methods of practice for particular subject areas.
As we shall see, the unresolved tensions between these assumptions and presumptions are responsible for the current state and outstanding questions of RLHF research.
This section does not set out to be a survey but rather interrelates core references to illustrate the modus operandi of RLHF and preference modeling.

To begin, all of the following operates on the assumptions that human preferences exist in any form, which emerged in early philosophical discussions, such as Aristotle's Topics, Book Three.

\begin{assumption}
Human preferences and goals exist.  
\end{assumption}

\subsection{Specifying objectives: from logic of utility to reward functions}

The optimization of RLHF explicitly relies only on reward models. 
In order to use rewards as an optimization target, RLHF presupposes the convergence of ideas from preferences, rewards, and costs.
Models of preference, reward functions, and cost landscapes all are tools used by different fields to describe a notion of relative goodness of specific actions and/or states in the domain. 
The history of these three framings dates back to the origins of probability theory and decision theory.
In 1662, \textit{The Port Royal Logic} introduced the notion of decision making quality~\citep{arnauld1861port}:
\begin{quote}
    To judge what one must do to obtain a good or avoid an evil, it is necessary to consider not only the good and evil in itself, but also the probability that it happens or does not happen. 
\end{quote}
This theory has developed along with modern scientific thinking, starting with Bentham's utilitarian \textit{Hedonic Calculus}, arguing that everything in life could be weighed~\citep{bentham1823hedonic}.
The first quantitative application of these ideas emerged in 1931 with Ramsey's \textit{Truth and Probability}~\citep{ramsey2016truth}. 

\begin{assumption}
\label{ass:1}
Any and all preferences and goals can be quantified and measured.
\end{assumption}

Since these works, quantifying, measuring, and influencing human preferences has been a lively topic in the social and behavioral sciences. 
These debates have rarely been settled on a theoretical level; rather, different subfields and branches of social science have reached internal consensus on methods and approaches to preference measurement even as they have specialized relative to each other, often developing their own distinct semantics in the process. 

A minority of economists posit that preferences, if they do exist, are prohibitively difficult to measure because people have preferences over their own preferences, as well as each others' preferences \citep{hirschman1984against}. 
In this view, which is not reflected in the RLHF process, individual preferences are always embedded within larger social relations, such that the accuracy of any preference model is contingent on the definition and context of the task. 
Some behavioral economists have even argued that preferences don't exist--they may be less an ontological statement of what people actually value than a methodological tool for indirectly capturing psychological predispositions, perceived behavioral norms and ethical duties, commitments to social order, or legal constraints~\citep{hadfield2014microfoundations}.
We address the links of this work to the Von Neumann-Morgenstern (VNM) utility theorem and countering impossibility theorems around quantifying preference in Sec.~\ref{sec:origins3}.


On the other hand, the reinforcement learning optimization methods used today are conceptualized around optimizing estimates of reward-to-go in a trial~\citep{sutton2018reinforcement}, which combines the notion of reward with multi-step optimization. 
The term \textit{reward} emerged from the study of operant conditioning, animal behavior, and the \textit{Law of Effect}~\citep{thorndike1927law,skinner2019behavior}, where a reward is a scale of ``how good an action is'' (higher means better).

Reward-to-go follows the notion of utility, which is a measure of rationality~\citep{briggs2014normative}, modified to measure or predict the reward coming in a future time window.
In the context of the mathematical tools used for reinforcement learning, utility-to-go was invented in control theory, specifically in the context of analog circuits in 1960~\citep{widrow1960adaptive}.
These methods are designed around systems with clear definitions of optimality, or numerical representations of goals of an agent. 
Reinforcement learning systems are well known for their development with a discount factor, a compounding multiplicative factor, $\gamma \in [0,1]$, for re-weighting future rewards.
Both the original optimal control systems stand and early algorithms for reward stand in heavy contrast to reward models that aggregate multimodal preferences.
Specifically, RL systems expect rewards to behave in a specific manner, quoting~\citet{singh2009rewards}:
\begin{quote}
Rewards in an RL system correspond to primary rewards, i.e., rewards that in animals have been hard-wired by the evolutionary process due to their relevance to reproductive success.
... Further, RL systems that form value functions, ... effectively create conditioned or secondary reward processes
whereby predictors of primary rewards act as rewards themselves...
The result is that the local landscape of a value function gives direction to the system’s
preferred behavior: decisions are made to cause transitions to higher-valued states.
A close parallel can be drawn between the gradient of a value function and incentive motivation~\citep{mcclure2003computational}.
\end{quote}
To summarize, rewards are used in RL systems as a signal to tune behavior towards clearly defined goals.
The core thesis is that an learning algorithm's performance is closely coupled with notions of \textit{expected fitness}, which permeates the popular view that RL methods are \textit{agents} that act in environments.
This view is linked to the development of reinforcement learning technology, exemplified by claims of the general usefulness of the reward formulation~\citep{silver2021reward}, but is in conflict when many individual desires are reduced to a single function.

\begin{assumption}
\label{ass:2}
Increasing the score of raw reward measurements corresponds to better behaviors (or value functions learned under invariant reward transformation~\citep{ng1999policy}).
\end{assumption}

\subsection{Implementing optimal utility}
Modern reinforcement learning methods depend strongly on the Bellman equation~\citep{bellman1957markovian, howard1960dynamic} to recursively compute estimates of reward-to-go, derived within closed environments that can be modeled as a Markov Decision Process (MDP)~\citep{sutton2018reinforcement}.
These origins of RL are inspired by dynamic programming methods are were developed solely as optimal control techniques (i.e. RL did not yet exist).
The MDP formulation provides theoretical guarantees of performance by structuring the environment as one with a non-changing distribution of state-actions.
\begin{assumption}
\label{ass:3}
Optimal solutions to reward maximization problems exist.
\end{assumption}

The term reinforcement, coming from the psychology literature, became intertwined with modern methods afterwards in the 1960s as \textit{reinforcement learning}~\citep{MENDEL1970287, waltz1965}.
Early work reinforcement learning utilized supervised learning of reward signals to solve tasks.
Work from Harry Klopf reintroduced the notion of trial-and-error learning~\citep{klopf1972brain}, which is crucial to success the field saw in the 1980s and on.

Modern RL algorithms build within this formulation of RL as a tool to find optimal behaviors with trial-and-error, but under looser conditions.
The notion of temporal-difference (TD) learning was developed to aid agents in both the credit assignment and data collection problems, by directly updating the policy as new data was collected~\citep{sutton1988learning}, a concept first applied successfully to Backgammon~\citep{tesauro1995temporal} (rather than updating from a large dataset of cumulative experience, which could be outdated via erroneous past value predictions).
The method Q-learning, the basis for many modern forms of RL, learns a model via the Bellman equation that dictates how useful every state-action pair is with a TD update~\citep{watkins1992q}\footnote{The term ``Q'' is used in Q-learning to refer to a technical concept the Q-function, which maps from any state-action to a scalar estimate of future reward.
A value-function maps from states to this same estimate.}.
Crucially, these notions of provable usefulness through utility have only been demonstrated for domains cast as MDPs or addressed in tasks with a single closed-form reward function, such as prominent success in games with deep learning (DQN)~\citep{mnih2013playing}.
Deep learning allowed the methods to ingest more data and work in high dimensionality environments.

As the methods became more general and successful, most prominent developments before ChatGPT had remained motivated within the context of adaptive control, where reward and cost functions have a finite notion of success~\citep{golnaraghi2017automatic}, e.g. a minimum energy consumption across an episode in a physical system.
Prominent examples include further success in games~\citep{silver2017mastering}, controlling complex dynamic systems such as nuclear fusion reactors~\citep{degrave2022magnetic}, and controlling rapid robotic systems~\citep{Kaufmann2023fpv}.
Most reward or cost functions can return an explicit optimal behavior, whereas models of human preferences cannot.
\begin{presumption}
\label{pres:1}
Optimal solutions can be achieved with finite data in complex environments.
\end{presumption}

Given the successes of deep RL, it is worth  noting that the mechanistic understanding of how the methods succeed is not well documented. 
The field is prone to mistakes of statistical analysis as the methods for evaluation grow more complex~\citep{agarwal2021deep}.
In addition, there is little mention of the subfield of inverse reinforcement learning (IRL) in the literature of RLHF. 
IRL is the problem of learning a reward function based on an agent's behavior~\citep{ng2000algorithms} and highly related to learning a reward model.
This primarily reflects the engineering path by which a stable approach to performing RLHF emerged, and motivates further investment and comparison to IRL methods to scale them to the complexity of open-ended conversations.

\subsection{Steering preferences}
\label{sec:origins3}

The context in which reinforcement learning was designed means that rewards and costs are assumed to be stable and determinative. Both rewards and costs are expected to be functions, such that if the agent is in a specific state-action pair, then it will be returned a certain value. 
As we move into preferences, this is no longer the case, as human preferences constantly drift temporally throughout their experiences. 
The overloading of the term ``value'' within these two contexts complicates the literature of RLHF that is built on the numerical value updates in Bellman equations with the very different notion of what is a human value, which often refers to moral or ethical principles, but is not well defined in technical literature.
An example of where this tension can be seen is how reward models are attempting to map from the text on the screen to a scalar signal, but in reality, dynamics not captured in the problem specification influence the true decision~\citep{salha2011aesthetics,gilbert2022choices}, such as preference shift when labeling many examples sequentially and assuming they are independent. 
Therein, modeling preferences is at best compressing a multi-reward environment to a single function representation.

In theory, the Von Neumann-Morgenstern (VNM) utility theorem gives the designer license to construct such functions, because it ties together the foundations of decision theory under uncertainty, preference theory, and abstract utility functions~\citep{von1947theory}; together, these ideas allow preferences to be modeled in terms of expected value to some individual agent.
The MDP formulation used in most RL research has been shown in theory to be modifiable to accommodate the VNM theorem~\citep{pitis2019rethinking}, but this is rarely used in practice.
Specifically, the Markovian formulation is limited in its expressivity~\citep{pitis2023consistent} and the transition to partially-observed processes, which is needed for language, further challenges the precision of problem specification~\citep{abel2021expressivity}.

However, the VNM utility theorem also invokes a number of assumptions about the nature of preferences and the environment where preferences are being measured that are challenged in teh context of RLHF.
Human-computer interaction (HCI) researchers, for example, have emphasized that any numerical model of preference may not capture all the relevant preferences of a scenario. 
For example, how choices are displayed visually influences people's preferences~\citep{salha2011aesthetics}. 
This means that representing preferences may be secondary to how that representation is integrated within a tool available for people to use. 
Work from development economics echoes this notion, showing that theories of revealed preferences may just recapitulate \textit{Hume's guillotine} (you can't extract an ``ought'' from an ``is''), and in particular the difference between choice (what do I want?) and preference (is X better than Y?)~\citep{sen1973behaviour}.

On a mathematical level, well-known impossibility theorems in social choice theory show that not all fairness criteria can be simultaneously met via a given preference optimization technique~\citep{arrow1950difficulty,maskin2014arrow}. 
Theoretical challenges to these theorems exist, for example by assuming that interpersonal comparison of utility is viable~\citep{harsanyi1977rule}.
That assumption has inspired a rich line of work in AI safety and value alignment inspired by the principal-agent problem in behavioral economics \citep{hadfield2016cooperative}, and may even include multiple principals \citep{fickinger2020multi}.
However, the resulting utility functions may come into tension with desiderata for corrigibility, i.e. an AI system's capacity to cooperate with what its creators regard as corrective interventions \citep{soares2015corrigibility}.
Philosophers have also highlighted that preferences change over time, raising fundamental questions about personal experiences, the nature of human decision-making, and distinct contexts~\citep{pettigrew2019choosing}.
These conflicts around the preference aggregation across people, places, or diverse situations is central to modern RLHF dataset engineering.

In practice, the VNM utility theorem ignores the possibility that preferences are also uncertain because of the inherently dynamic and indeterminate nature of value---human decisions are shaped by biology, psychology, culture, and agency in ways that influence their preferences, for reasons that do not apply to a perfectly rational agent. 
As a result, there are a variety of paths through which theoretical assumptions diverge in practice:
\begin{itemize}
\item measured preferences may not be transitive or comparable with each other as the environment where they are measured is made more complex;
\item 
proxy measurements may be derived from implicit data (page view time, closing tab, repeating question to language model), without interrogating how the measurements may interact with the domain they're collected in via future training and deployment of the model;

\item the number and presentation of input sources may vary the results, e.g. allowing respondents to choose between more than two options, or taking in inputs from the same user at multiple times or in multiple contexts; 
\item relatively low accuracy across respondents in RLHF training data, which may mask differences in context between users that the preference model can aggregate or optimize without resolving.
\end{itemize}

\begin{presumption}
\label{pres:2}
The temporal- and context-shifting of user preferences does not mitigate the effectiveness of reward functions or notions of optimal utility as an optimization target.
\end{presumption}

\begin{figure}[t]
    \centering
    \includegraphics[width=\linewidth]{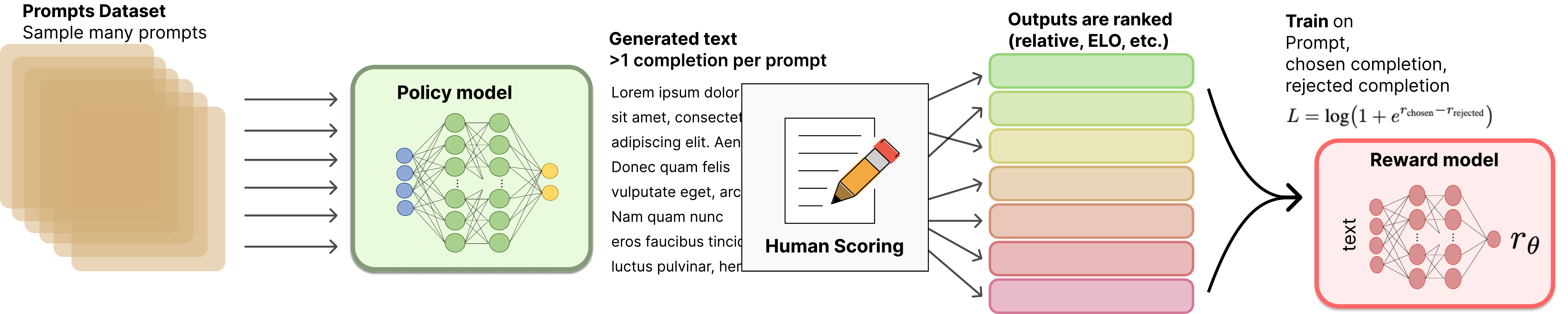}
    \caption{Illustrating the training process of an RLHF language reward model.
    A language model generates text from a distribution of prompts to optimize human preferences for, then humans assign a preference order to them (normally pairwise preferences).
    These preferences are used to train a score function via a contrastive loss shown in Eq.~\ref{eq:pm_loss}.}
    \label{fig:rlhf2}
\end{figure}

\section{Background}
\label{sec:back}
We continue to use \textit{assumptions} of the literature, grounded in theoretical backing of a subject area, and \textit{presumptions}, which are commonly accepted methods of practice, to identify blind spots and open questions in reward modeling.

\subsection{Reward models of human preferences}
\paragraph{Data collection}
Reward models are trained with human preference data collected over a distribution of graphical user interfaces.
The most common task is to give the model a \textit{prompt}, i.e a request or instruction, and rate the \textit{completion}, or answer.
Implementation methods vary; some collect rankings from groups of responses~\citep{ziegler2019fine}, others collect scores and rankings of a group of candidate responses~\citep{ouyang2022training} (scores of 1-5 for 6+ outputs then ranking all), yet others present a choice between a pair of responses~\citep{bai2022training} (choose best response between two options), and more remain~\citep{wu2023fine}.
Pairwise preferences is often described as the base method~\citep{christiano2017deep}, derived from the Bradley-Terry model~\citep{BradleyTerry}.
The workers employed are generally given detailed instructions on which styles, occurrences, or values to prioritize in their labels.
This data is collected from groups of individuals and sometimes calibrated across per-labelor distributions, but that decision making process is not always documented or shared.

In each batch of rankings for a prompt-completion pair, the pairs are often \textit{binarized} into a chosen element and rejected element (rather than using the corresponding delta as a training objective) to create easy-to-optimize training data.
This binarization of chosen and rejected is aggregated independently to measurements of the difficulty of the prompt or mean quality of the responses.
\begin{presumption}
\label{pres:6}
Pairwise preferences can suitably perform as a basis of human values.
\end{presumption}

\paragraph{Model training}
The reward models trained for RLHF are most often trained as classifiers between a chosen and rejected completion to a prompt before optimizing with RL where they return a scalar value for each piece of text.

The loss function of reward models is formulated as a difference between the score for each binarized preference. 
Given two options for a completion $y$ from a prompt $x$, and the scores they obtain a scalar output $r$ from a value head of a language model, the loss for the reward model is as follows~\citep{askell2021general, ouyang2022training}:
\begin{equation}
L = \text{log}\big( 1+e^{r_\text{chosen} - r_\text{rejected}} \big)
\label{eq:pm_loss}
\end{equation}
The loss function is designed to increase the distance between the two samples, where variations exist including losses of 4 samples rather than a pair~\citep{ziegler2019fine}, updating the model with batches of pairwise labels on a given prompt~\citep{ouyang2022training}, or optimizing based on the margin between $r_\text{chosen}$ and $r_\text{rejected}$~\citep{touvron2023llama}.

There are multiple popular ways for adapting base language models into reward models.
In earlier work, a value head was appended to the model to predict these rewards.
A value head can be as simple as adding a linear layer with one output node after the attention blocks of a Transformer model.
Today, some work includes an entire copy of the base language model with modified output layers for predicting rewards.
Other reward models are trained by removing layers from a base model and/or freezing original model parameters during training.
The reward model training component of RLHF is shown in Fig.~\ref{fig:rlhf2}.
\begin{presumption}
\label{pres:3}
Multiple user preferences are successfully represented in training one model by aggregating and comparing individual utilities.
\end{presumption}
\textit{Note: Presumption 4 is the convergence of both training practices and motivation of preference models, as we addressed in Sec.~\ref{sec:origins3} where economists disagree as to whether preferences can be aggregated.
As we discuss, some economists construe this aggregation of preferences as an assumption core to the field, but the practice in RLHF is looser in its rigor due to its evolution from the MDP to the partially observed setting.
}
\begin{presumption}
\label{pres:4}
The only preferences embedded in the model are from the specifically collected training data.
\end{presumption}

\subsection{Reinforcement Learning on Language}
Language generation optimized via reinforcement learning, which RLHF is a version of, can be formalized as a 
partially observable Markov decision process (POMDP)~\citep{spaan2012partially}.
We define a POMDP $\mathcal{M}$ at a per-token level with $\mathcal{M}=(\mathcal{S}, \mathcal{A}, \mathcal{O}, \mathcal{T}, \mathcal{Z}, \mu_0, \mathcal{R}, \gamma) $. 
Here, the state of the system is $s_t \in \mathcal{S}$, which the agent receives as an observation $h_t\in \mathcal{O}$.
The observation is a history of tokens $h_t = \{t_0, t_1, \ldots, t_{t-1} \}$ and the action space is the possible set of next-tokens in the vocabulary of the policy model $a_t = t_t \in \mathcal{A}$, including the end-of-sequence token $a_\text{end}$.
As in a traditional MPD, $\mathcal{T}$ is the transition function $\mathcal{T}(\cdot|s_t, a_t)$.

The goal of the RLHF process is to learn a policy that maps $\pi : \mathcal{O}\mapsto\mathcal{P}(\mathcal{A})$.
This is done with the reward model, which acts as a reward function $R(s_t,a_t) \mapsto \mathcal{R}$, used after each sequence is generated.
The full sequence, until end-of-sequence token $a_\text{end}$, is called the action and used to get a scalar reward $r_t$ from the reward model.

The rewards for the batch are used for the RL update, where a popular algorithm is Proximal Policy Optimization~\citep{schulman2017proximal}.
In RLHF, the discount factor of reward is set to 1 and no further actions are taken until the end of a sequence, casting the problem as contextual bandits.
\begin{presumption}
\label{pres:5}
User preferences are extracted uniformly via the RLHF process.
\end{presumption}
\section{Questions}
\label{sec:questions}

In this section, we propose a series of investigations to disentangle the diverse assumptions around using large models of human preferences built on LLMs within the RLHF framework.
We focus on the core sections of a machine learning solution: data, model, and the optimization choices that dictate which priorities are encoded in contemporary reward models.
The questions are connected to one or more assumptions or presumptions presented in Sec.~\ref{sec:origins} and Sec.~\ref{sec:back}.

\subsection{Model Questions}
The reward models of human preference used in RLHF today are built on base, generative LLMs, so issues from the full design and training process of these models applies to reward models.
This includes the wide literature on potential sources of harm from LLMs, including a large swath of social biases, such as anti-Muslim bias~\citep{liang2021towards, abid2021persistent}.
Those pretraining LLMs should reckon with the design choices of the base model including environmental costs, training datasets, stakeholder involvement, and more~\citep{bender2021dangers}.
Base model capability distributions should be tested against, but often are not for engineering and competitive reasons.
Given this, two sets of evaluations should be included in the RLHF process: 
\begin{itemize}
\item 
\textbf{Base model biases} (for \textbf{Presumption}~\ref{pres:4}):
How do different base LLMs used in the reward model training process influence all aspects of reward models? 
How does this impact training numerical stability, efficiency, training or test loss, and more?
Is it possible that many design choices in RLHF are based on a static model and could be solved with a change of base model, as opposed to increased data and optimization?
This links to all the questions we ask in the next subsection -- ultimately, how does the base model perform as an independent variable?
\item 
\textbf{Sequential model evaluation in RLHF} (for \textbf{Presumption}~\ref{pres:4}, \ref{pres:5}):
When a reward model is trained, it is making subtle changes to the parameters of a LLM. 
This is done by appending parameters in parallel, i.e. re-routing the model, so the text that the reward model produces could be observed. 
How do the biases of a pretrained base model, a downstream RLHF model, and the proxy of an intermediate reward model showcase how RLHF modifies representations of LLMs?
Given the common implementation technique of appending a value head to a base LLM and fine-tuning to construct a reward model, the reward models can still generate text that can indicate if training is operating as intended.
The base model could also impact which data from the preferences dataset is learned during the RLHF process.
\end{itemize}

\subsection{Data Questions}
The data used to enable RLHF is often curated by multiple stakeholders in a combination of paid employment and consumer usage.
This data, representing a preference between two pieces of text in an individual instance, is capturing a broad and diverse function via extremely limited interactions.
Given that the data is sparse in count relative to the complexity it begins to represent, more questions should be openly shared about its curation and impacts.

Currently, datasets for the most popular LLMs are being generated by professional work-forces. 
This opens up many questions around who is creating the data and how the context of their workplace informs it, including:
\begin{itemize}
\item \textbf{Data collection contexts} (for \textbf{Presumption}~\ref{pres:2}):
Can data involving preferences collected in a professional setting mirror the intent of researchers designing an experiment or provide suitable transfer to downstream users? 
How does this compare to volunteer workers? 
How does context inform preferences, how does this data impact a downstream model, how can the impact of a user interface be measured in data?
How does repetitive labeling of preference data shift one's preferences? 
Do professional crowd-workers, instructed to follow a set of preferences, follow the instructions or their innate values? 
\item \textbf{Type of feedback} (for \textbf{Presumption}~\ref{pres:6}): Does the default operating method of RLHF, pairwise preferences capture preferences in its intended form? 
Can comparisons in RLHF across the same data be made with the default comparisons versus advanced multi-axes feedback mechanisms~\citep{wu2023fine}?
What types of comparisons would reflect how humans communicate preferences in text?
\item \textbf{Population demographics} (for \textbf{Presumption}~\ref{pres:3}): Who is completing the data? 
Is a diverse population maintained?
How does a lack of diversity emerge as measurable impacts on the model?
What is a minimum number of people required to suitably represent a given population?
How are instances of preference annotator disagreement treated -- as a source of noise, or a signal?
\end{itemize}

\subsection{Optimization questions}
The most elusive and potentially insightful questions with respect to reward models and preference data emerge where preferences are extracted by reinforcement learning optimizers and distilled into the LLM.
In this vein, further questions should be asked as to whether the outcomes of the optimization match the intended design of the process.
Given the complexity of LLMs and the social values of using human data to empower technology, clear attribution is needed for the role of the preference data in the optimization.
Questions that can be used to investigate these issues include:
\begin{itemize}
    \item \textbf{RL optimization of reward model} (for \textbf{Assumption}~\ref{ass:1},~\ref{ass:2},~\ref{ass:3}, \textbf{Presumption}~\ref{pres:1}): at a technical level, is the reward signal from the reward model for preference maximized? 
    Should it be maximized?
    Does this reflect changes in line with the preference data (e.g. is this achieved at the same point where evaluation accuracy peaked). 
    In the RL literature, sharing learning curves is common practice, yet in RLHF it is often omitted, with just the final test set accuracy included. 
    Correlating these common metrics is important to understanding the role of the reward model to the process.
    \item \textbf{Qualitative alignment} (for \textbf{Assumption}~\ref{ass:1}, \textbf{Presumption}~\ref{pres:2},~\ref{pres:3},~\ref{pres:5}): Given the RLHF data process starts with written instructions to data-workers, checks should be done to identify if the model changes align with the intended goals of stated preferences.
    A suitable method for continued deployment to iteratively refine training and deployment goals would be documentation for RL systems, \textit{Reward Reports}~\citep{gilbert2022reward}.
    Beyond this, feedback from annotators, data analysis, and more can quickly provide feedback onto the impacts of RLHF relative to human inputs and intentions.
    \item \textbf{Weighing preferences} (for \textbf{Presumption}~\ref{pres:5}): Does and should the model uniformly extract information from the dataset? 
    How are different scenarios in the dataset equated (e.g. instructions around healthcare being treated the same as food choices)? 
    Should different reward functions derived in preference literature be used to maintain an inductive bias over preferences, such as Plackett-Luce~\citep{cheng2010label}.
    Recent literature shows that aggregate agreement between humans and the reward models varies from approximately 60 to 75\%~\citep{stiennon2020learning, ouyang2022training, bai2022training}, should certain examples be prioritized given an non-uniform distribution of tasks and annotators? 
    Or, can these distributions be broken down by commonly accepted categories such as reasoning, generation, etc.? 
    Such reduction to the mean across groups can squash multi-modal beliefs, especially with underrepresented groups~\citep{prabhakaran2021releasing, feffer2023moral}.
\end{itemize}


\section{Solutions}
\label{sec:sol}

\paragraph{Evaluation of reward model capabilities}
The most common evaluation technique for reward models is to hold out data from the training set and measure the final model's agreement with the data, but this does not control for the complex nature of the dataset or encourage cross-intuition comparisons.
The evaluation score is unlikely to illuminate any conflicts in the data that arise from aggregating over inputs from multiple individuals or contexts. 
This necessitates building more controlled examples to interpret whether the  objective of differentiating better from worse text is achieved. 
Taking inspiration from recent advancements in evaluating instruction- and chat-tuned LLMs~\citep{zheng2023judging}, where the existing academic benchmarks from the natural language processing literature were incomplete for evaluating modern models like GPT4~\citep{chang2023survey}, the community should evaluate reward models in a manner analogous to their use.
One specific example exists in recent work, a benchmark to evaluate reward model consistency~\citep{shen2023trickle}, as its ability to maintain consistent scores over changes to text that do not alter meaning, but many more tools are needed to cover the many use cases of RLHF models.  
Similar to MT Bench~\citep{zheng2023judging}, a benchmark for LLM capabilities over multi-turn conversations, we can construct representative examples across a variety of categories for the model (e.g. reasoning, generation ability, factuality, etc.) to evaluate the reward signal.

\paragraph{Evaluation of reward model safety}
The primary means for evaluating the safety, toxicity, and potential harms of LLMs is a process known as red-teaming, which entails focused attempts to prompt the language model into a broad array of harmful behaviors~\citep{ganguli2022red}. 
Red-teaming for generative models is primarily separated into two axes, attack vectors and harms.
The notion of red-teaming for reward models is simpler and less interactive, as the output of the model is a score rather than text to read and comprehend.
To do so, the score of reward models across a series of text documents or snippets that have the potential for harm should be compared to `neutral' texts.
Second, reward models should be evaluated for adversarial input strings, e.g. sequences of text that produce unexpectedly high or low scores, which could indicate downstream exploitation during the RL step.
As a starting point, practitioners could consider the efforts to identify adversarial prompts that escalate toxicity (but need not necessarily do so) in the evaluation of LLM generations, for instance as in done in the RealToxicityPrompts dataset \citep{gehman2020realtoxicityprompts}.
Today, a third-party researcher can play with a closed model and find an adversarial attack even if no red-teaming protocols are shared about the system. 
The same is not true for a private reward model, which may be re-used in the future to train the next model. 

\paragraph{Sociotechnical specification of preference}
The current manner by which reward models are used in RLHF follows the logic of the following statement: \textit{the reward model is deemed good when it is good for the downstream task}.
Given the complex intellectual origins of quantifying and optimizing preferences, specific recommendations should be made around what information a model of human preferences should and should not encode.
Doing so may reduce the performance of models in the short term, but increases the ability for multi-stakeholder engagement and reduction of harms with the development of LLMs.

\section{Discussions}
\label{sec:discussion}

\paragraph{Downstream use of models}
The final manner by which the assumptions of RLHF are tested is in the increasing alignment of LLM training with the intentions of downstream users, as is the case for ChatGPT~\citep{schulman2023proxy}.
LLMs are technologies used in consumer applications, giving different context and broader use-cases than the limited training data.
The developers of RLHF should consider how end user encounters with their model may be different than the preferences collected for training, such as the shift from primarily English training data to multilingual users. 
Given the direct manner by which human preferences are encoded in models trained with RLHF, users of these models should be presented with the intended preferences of the models for transparency.
A potential improved solution is to let users choose a model matching their preferences, but this comes with many technical challenges.
Emerging examples of LLM `hub' interfaces that offer users a menu of LLMs to choose from (like the open-source GPT4All\footnote{\url{https://gpt4all.io/index.html}}, or Meta's new AI companion character suite\footnote{\url{https://about.fb.com/news/2023/09/introducing-ai-powered-assistants-characters-and-creative-tools}}) offer hope that such approaches might in-fact be tractable and/or accessible to users.

\paragraph{Synthetic preference data}
Recent work on Reinforcement Learning from AI Feedback (RLAIF) uses LLMs in addition to humans to provide critiques of other LLM generations for the RLHF process~\citep{bai2022constitutional, lee2023rlaif, wang2023shepherd}. 
Further, these methods are referenced as methods for removing the human bottleneck from alignment~\citep{superalignment} without existing literature on how methods such as RLHF and reward models succeed in capturing values.
Despite the practical attractiveness due to the cost of acquiring human data, synthetic preference data presents different challenges.
Preliminary findings have indicated that synthetic supervision data for LLMs can lead to generation instability and lack of robustness~\citep{alemohammadSelfConsuming2023,gudibandeFalse2023,shumailovCurse2023a}, to say nothing of the potential for further shifting preferences away from the original human data.

\paragraph{Direct preference optimization}
Recent work has shown that direct optimization can be used to extract information from human preference data~\citep{rafailov2023direct}, rather than using an intermediate reward model.
The removal of the reward model would make the encoding of values more opaque in the RLHF process, yet at the same time potentially reduce problem misspecification and exploitation by reducing the number of training steps.
The questions suggested in this work would still apply, but the methods for auditing them will be different without a distinct reward model.

\section{Conclusion}
Given the importance of RLHF to the deployment and integration of state-of-the-art LLMs, a complete understanding of the motivation, intellectual foundations, and implementation presumptions of reward models is crucial to the safe and equitable rollout of this technology.
With the evolution from utility quantification to reinforcement learning to maximizing average preference scores, we have shown that the assumptions at each step are not in-step with how the technology is used in contemporary settings.
This paper details the intellectual basis of learning reward models, and argues that more information about existing reward models should be shared with the public via open or partially-open releases.
With more development and sharing of reward models, proper evaluation techniques of the models can be designed, which will lead to further technological developments around desired properties such as user tunability, uncertainty quantification, etc.
By doing this work, we can better understand the distinct challenges of \textit{learning} models from incomplete data of human preferences and then \textit{optimizing} those models to elicit specific behaviors.

\section*{Acknowledgements}
Thanks to Aaron Snoswell for providing edits and comments on this work.
Credit to Sasha Luccioni and Deep Ganguli for discussions and encouragement towards completing this paper.
Credit to Silviu Pitis for providing substantial feedback on the first version of the paper, improving the robustness and depth of the argument.
Thanks to Kristian Lum and Dylan Hadfield-Menell for substantial feedback on the early versions of this paper.

\bibliography{rlhf}

\end{document}